\documentclass[%
reprint,
superscriptaddress,
amsmath,amssymb,
aps,
pre,
longbibliography,
]{revtex4-2}

\usepackage{color}
\usepackage{textcomp}
\usepackage{gensymb}
\usepackage{graphicx}
\usepackage{dcolumn}
\usepackage{bm}
\usepackage{hyperref}
\usepackage{epstopdf}
\usepackage[export]{adjustbox}
\usepackage{float}

\begin{document}
\preprint{APS/123-QED}

\title{Capillary Thinning of Elastic and Viscoelastic Threads: from Elastocapillarity to Phase Separation}

\author{H. V. M. Kibbelaar}
 \affiliation{Van der Waals-Zeeman Institute, Institute of Physics, University of Amsterdam, 1098 XH Amsterdam, The Netherlands.}
\author{A. Deblais}
 \affiliation{Van der Waals-Zeeman Institute, Institute of Physics, University of Amsterdam, 1098 XH Amsterdam, The Netherlands.}
 \affiliation{Unilever Innovation Centre Wageningen, Bronland 14, 6708 WH Wageningen, The Netherlands.}
\author{F. Burla }
\affiliation{AMOLF, Department of Living Matter, 1098 XG Amsterdam, The Netherlands.}
\author{G. H. Koenderink}
\affiliation{AMOLF, Department of Living Matter, 1098 XG Amsterdam, The Netherlands.}
\affiliation{Department of Bionanoscience, Kavli Institute of Nanoscience, Delft University of Technology, 2629 HZ Delft, The Netherlands.}
\author{K. P. Velikov}
\affiliation{Van der Waals-Zeeman Institute, Institute of Physics, University of Amsterdam, 1098 XH Amsterdam, The Netherlands.}
\affiliation{Unilever Innovation Centre Wageningen, Bronland 14, 6708 WH Wageningen, The Netherlands.}
\author{D. Bonn}
\affiliation{Van der Waals-Zeeman Institute, Institute of Physics, University of Amsterdam, 1098 XH Amsterdam, The Netherlands.}

\date{\today}
             
\begin{abstract}
{The formation and destabilisation of viscoelastic filaments are of importance in many industrial and biological processes. Filament instabilities have been observed for viscoelastic fluids but recently also for soft elastic solids. In this work, we address the central question how to connect the dynamical behavior of viscoelastic liquids to that of soft elastic solids. We take advantage of a biopolymer material whose viscoelastic properties can be tuned over a very large range by its pH, and study the destabilization and ensuing instabilities in uniaxial extensional deformation. In agreement with very recent theory, we find that the interface shapes dictated by the instabilities converge to an identical similarity solution for low-viscosity viscoelastic fluids and highly elastic gels. We thereby bridge the gap between very fluid and strongly elastic materials. In addition, we provide direct evidence that at late times an additional filament instability occurs due to a dynamical phase separation.}   
\end{abstract}
\pacs{Valid PACS appear here}
\keywords{Fluid dynamics, Soft Matter}
\maketitle

Viscoelastic fluids exhibit an array of properties that differentiate them from Newtonian fluids like water \cite{Larson1999,Amarouchene2001,Mckinley2005}. Striking differences in flow behavior are observed due to the presence of mesoscopic constituents such as polymers that introduce an elasticity into an otherwise viscous system. Especially the formation of viscoelastic filaments has attracted much attention. The obvious importance of viscoelastic filaments for fibre spinning \cite{Naraghi2007,Keshavarz2020} has become a benchmark problem for testing viscoelastic fluid mechanics \cite{Anna2001,Mckinley2002,Furbank2004,Suryo2006,Smith2010,Miskin2012,Huisman2012}. When pushing a polymeric liquid out of a syringe, instead of breaking off at the orifice by a Rayleigh-Plateau instability to form a drop as a Newtonian fluid does \cite{Eggers1997,Deblais2018b}, long and slender filaments form that are very long-lived. These subsequently undergo spectacular beads-on-a-string (BOAS) \cite{Goldin1969,Bazilevskii1981,Entov1984,Wagner2005,Bhat2010,Clasen2006} and blistering \cite{Oliveira2005,Sattler2008,Sattler2012} instabilities while undergoing capillary thinning. Such BOAS instabilities can be viewed as the viscoelastic equivalent of the Rayleigh instability. However, due to the elasticity of the filaments these do not break up until very late times. Recently, instabilities strikingly similar to the BOAS structure have been observed on filaments made of very soft elastic gels \cite{Mora2010} that do not break up at all. This poses the question how to connect the dynamical behavior of viscoelastic liquids to that of soft elastic solids. 

\begin{figure}[b]
\includegraphics[scale=1]{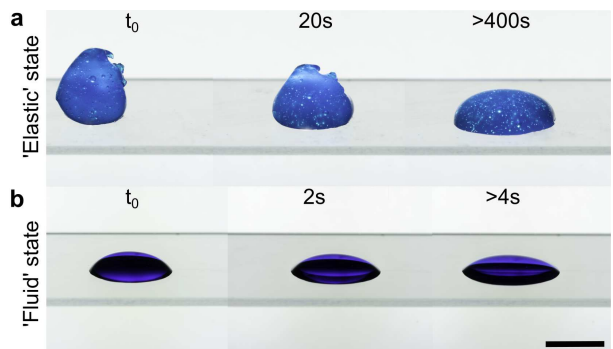}
\caption{Timelapse photographs of blue coloured droplets of HA (3 mL volume), on a glass substrate, prepared at pH 2.5 representing the elastic state (a, upper panel) and at pH 7 representing the fluid state (b, lower panel). The fluid solution spreads like a liquid, while the elastic sample only flows on a time scale of minutes. The scale represents 15 mm.}
\label{F1}
\end{figure}

\begin{figure*}[t]
\includegraphics[scale=1]{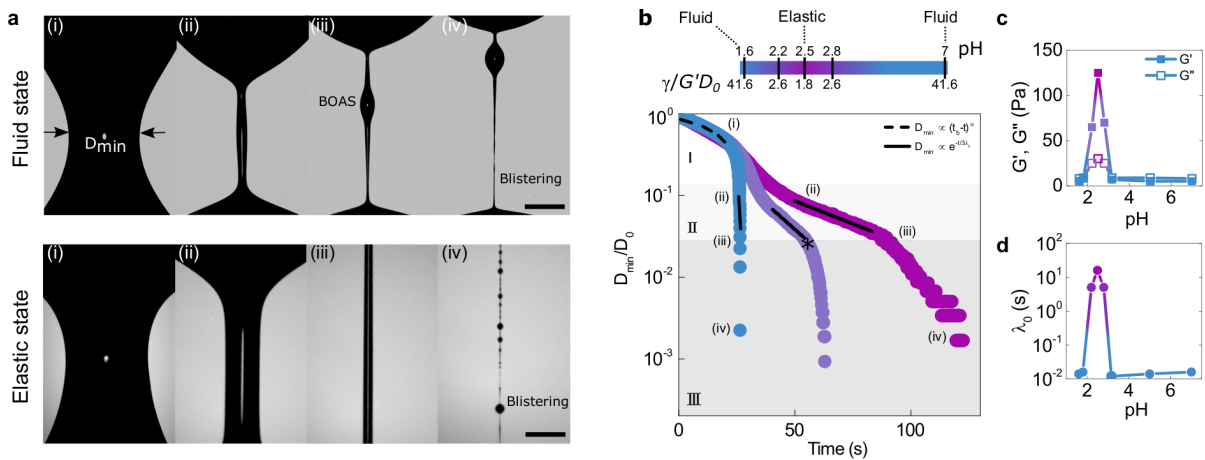}
\caption{(a) Photographs of the neck breakup dynamics of HA in the fluid state (upper panel, pH 7) and in the elastic state (lower panel, pH 2.5). The scale represents 75 $\mu$m in the upper panel and 400 $\mu$m in the lower panel. (b) Minimum filament diameter normalized by the initial bridge size as a function of time for three different solutions at different elastocapillary numbers $\gamma/G'D_{0}$ given in the upper colored bar: blue for pH 1.6 and 7, violet for 2.2 and 2.8 and purple for 2.5. The numbers correspond to the photographs in (a). (c) The viscoelastic moduli from oscillatory shear measurements. (d) Relaxation times obtained from (b).}
\label{F2}
\end{figure*}

Very recent theory by Snoeijer \textit{et al.} \cite{Snoeijer2019} and Eggers \textit{et al.} \cite{Eggers2020} investigates this question in detail. In the generic models for the flow of viscoelastic polymeric liquids, one has to adopt a Lagrangian reference frame: the polymers that are being stretched move with the fluid and are stretched by velocity gradients in the moving fluid. For elasticity theory, on the other hand, the reference frame is Eulerian, and all the deformations are calculated in the laboratory frame of reference. These recent theories show that the two reference frames can be mapped onto each other (at least for some explicit models for polymeric fluids) by considering the relaxation time scale $\lambda$ of the polymer. When this relaxation time goes to infinity, the deformations calculated in the moving reference frame for polymeric fluids simplify to purely elastic (but non-linear) deformations in the elastic models. The most striking conclusion is that in practice the instability is identical between purely elastic and viscoelastic materials. For instance, for the pertinent example of the BOAS structure the prediction is that the interface shapes connecting the filament (the string) to the drops (the beads) has an identical shape \cite{Snoeijer2019,Eggers2020, Turkoz2018}. To test these proposals and investigate how the crossover between elastic and viscoelastic filaments happens we here investigate the capillary thinning of filaments of a biopolymer with a tunable elasticity, varying from an almost Newtonian liquid to a gel. 

We take advantage of the fact that the viscoelastic properties of the biopolymer Hyaluronic Acid (HA) can be tuned by means of its pH \cite{Giubertoni2019}, and study the effect of elasticity on filament instabilities. HA solutions were prepared following a frequently used protocol \cite{Giubertoni2019,Burla}. 1 wt.$\%$ HA solutions were prepared using Hyaluronic Acid sodium salt powder from Streptococcus Equi bacteria (1.5 \textendash 1.8 MDa, Sigma Aldrich) in distilled water, together with a fixed NaCl concentration (0.15 M) and HCl concentrations ranging from 0 to 50 mM to obtain pH values from 1.6 to 7. Samples were vortexed for a few seconds to ensure mixing of the ingredients and homogenized under modest rotation during a period of 5 days. As illustrated in Fig.~\ref{F1}, the viscoelastic behavior of HA solutions strongly depends on the pH. Samples prepared at pH values from 1.6-1.9 and 3-7 form a fluid state and spread like a liquid, while an elastic state forms at pH 2.5 that retains its shape and only deforms significantly on a time scale of minutes. Oscillatory shear rheology at 0.5 Hz and a strain amplitude of 0.5\%, well within the linear regime of HA (see Supp Fig.~1 \cite{Sup}), also shows that at pH 2.5 an increase of the storage modulus of more than two orders of magnitude occurs (see Fig.~\ref{F2}(c) and Supp Fig.~2 for experimental details \cite{Sup}). For both lower and higher pH values than 2.5, the viscoelastic gel becomes a viscoelastic liquid. In the remainder we use the term `elastic' state for the solution state at pH 2.5 and for the other pH `fluid' state. Such responsiveness of biopolymers is of relevance biologically and consequently observed in various biological systems such as curvature driven instabilities of lipid bilayer membranes \cite{Campelo2007}, crawling cells \cite{Stossel1993}, and plays a role in the fabrication of scaffolds in tissue engineering \cite{And2001,Zhao2011}, 3D cell cultures \cite{Chaudhuri2016a,McKinnon2014}, but also in electrospun nanofibers \cite{Naraghi2007,Ji2006}, and in 3D printing \cite{Truby2016}. In a number of these examples, one observes the formation of filaments which are prone to the capillary instabilities that we investigate here. 

To study the extensional thinning and destabilization of HA filaments for different pH values, we use a custom built filament stretching rheometer (see Supp. Fig.~3 for sketch of set-up and experimental details \cite{Sup}), similar to the one described in \cite{Louvet2014, Huisman2012}. A small sample of 40 $\mu$L fluid is initially placed between two circular end plates ($D_0 = 5$ mm and $L_0 = 2.5$ mm) which are moved apart at a slow and constant velocity of 0.1 mm/s until the bridge breaks due to surface tension (the low velocity does not introduce any additional velocity in the problem). The evolution of the liquid bridge is recorded with a fast camera (Phantom V7) allowing frame rates up to 10.000 frames per second. The camera is equipped with a microscope tube lens, with an objective up to 12x magnification (Navitar) and a spatial resolution of 3 $\mu$m per pixel. The setup is placed in a closed chamber (80$\%$ RH) to prevent evaporation during the measurements. 

Figure~\ref{F2}(a) shows typical photographs of the break up dynamics for the fluid state (blue, at pH 7) and the elastic state (purple, at pH 2.5) and the corresponding thread radius as a function of time in a semi logarithmic plot (b), where also an intermediate state is added at pH 2.8 and 2.2 (violet) to emphasize the tunability of HA (see Supp.~Fig.~4 \cite{Sup} for the thread radius as function of time for all pH for the corresponding break up pictures see Supp. Fig.~5 in \cite{Sup}). The observed filament thinning of the HA solutions with different pH can be separated into three regimes. We first observe a power-law regime (I) where the neck thinning is similar to that of a power law fluid following a visco-capillary balance \cite{Renardy2002,Renardy2004,Doshi2004,Suryo2006,Huisman2012}. The break up pictures in Fig.~\ref{F2}(a) designated by (i) correspond to this regime. This regime follows $D_{min} = (t_{b}-t)^{\alpha}$ \cite{Clasen2010}, indicated by the black dashed fit in Fig.~\ref{F2}(b), where $t_b$ refers to the breakup time of the filament and $\alpha$ depends on the rheology of the fluid. The $\alpha$ values corresponding to the fluid and elastic state are respectively 0.3 and 0.9, which also correspond to the power law exponents from the shear thinning rheology of the polymers (Supp. Fig.~6 and Supp. Note~1. for experimental details \cite{Sup}). 

\begin{figure}[t]
\includegraphics[scale=1]{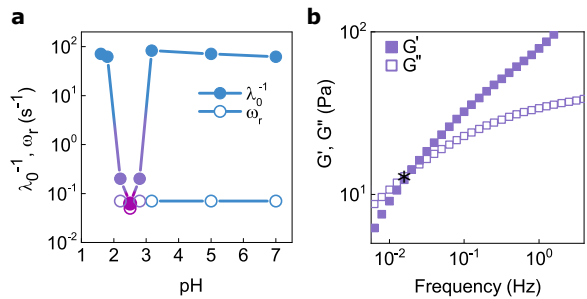}
\caption{(a) BOAS instabilities are observed when the growth rate of the Plateau-Rayleigh instability is smaller than the inverse characteristic time of the polymer (${\lambda}_{0}^{-1} > \omega_{R}$); for pH 2.5, no bead is formed (the color gradient is the same as in Fig.~\ref{F2}). (b) Viscoelastic moduli at pH 2.8 as function of frequency. The crossover point at very low frequencies (marked with an asterisk) indicates the timescale at which the BOAS instability appears, corresponding to the transition between regimes II and III in Fig.~\ref{F2}(b).}
\label{F3}
\end{figure}

The extensional rate is directly obtained from the evolution of the filament diameter as $\dot{{\varepsilon}} = \frac{-2}{D_{min}}\frac{dD_{min}}{dt}$, which are shown in the Supp. Fig.~7 \cite{Sup} for the fluid and elastic states. The extensional rate shows that in the powerlaw regime (I) the extension rate increases as the neck diameter decreases. This regime has been observed previously for strongly shear thinning but weakly elastic samples and reflects the modification of the Newtonian thinning regime by the shear thinning \cite{Louvet2014, Mckinley2002}. Subsequently, an exponential regime is reached where a long and slender cylindrical filament is formed (break up photographs indicated by (ii)): this is the classical elastocapillary regime (II) defined by the exponential thinning of the filament, $D_{min} \propto e^{-t/3\lambda_0}$ with $\lambda_{0}$ being the longest relaxation time of the polymer solution \cite{Anna2001}. The black straight lines in Fig.~\ref{F2}(b) indicate the exponential fit from which the relaxation times of the solutions are determined. For the low-elasticity sample called the fluid state there is also a short elastocapillary regime, as shown in the Supp. Fig.~3 \cite{Sup}. After the exponential regime II we observe an even steeper than exponential decay (regime III), eventually leading to the breakup. Within regime III instabilities occur: the BOAS structure is observed in the fluid state (photographs indicated by (iii)). For the elastic state, the cylinder remains symmetric and does not show BOAS instabilities. In addition, for all pH, just before the break up, a blistering pattern is observed (photographs indicated by (iv)) that is clearly very different from the BOAS. 

\begin{figure}[t]
\includegraphics[scale=1]{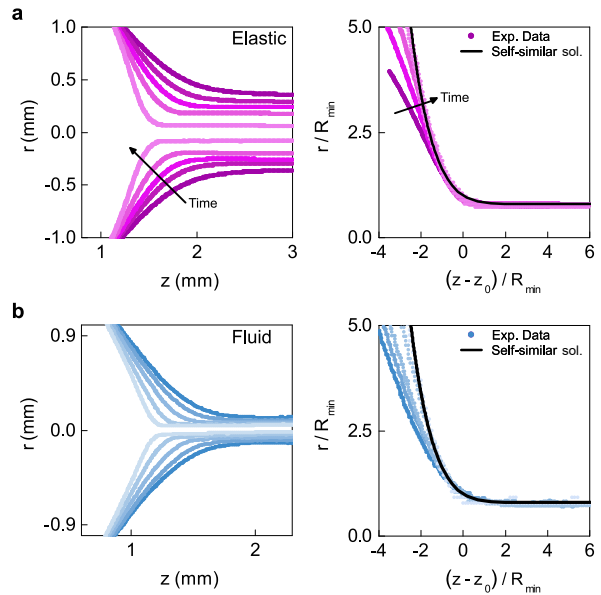}
\caption{Time evolution of the filament shapes in (a) the elastic (pH 2.5, $\gamma/G'D_{0}$=1.8), 12.5 s between each profile and (b) the fluid states (pH 7, $\gamma/G'D_{0}$=40.2), 10 ms between each profile. (Right) The same profiles but rescaled by the minimum neck radius $R_{min}$ and with $z_{0}$ the location for which the experimental profiles are collapsing. The universal self-similar solution is the black line.}
\label{F4}
\end{figure}

In Fig.~\ref{F2}(c) the storage $G'$ and loss $G''$ shear moduli are plotted as a function of pH and in Fig.~\ref{F2}(d) the corresponding relaxation times from the extensional rheology measurements on a logarithmic scale, as they cover several orders of magnitude. The longest relaxation time is observed for the elastic state at pH 2.5 and we find that the moduli follow the same trend as the relaxation times. 

We find that when the elastic modulus is the highest (at pH 2.5), the elastic stresses resist the formation of a bead. Comparing to Mora \textit{et al.} \cite{Mora2010}, their soft solids show instabilities when the ratio of the elastocapillary length to the cylinder diameter is larger than 6 ($\gamma/G'D_{0} > 6$). In our system, when we calculate the elastocapillary length for the elastic state at pH 2.5 at a finite frequency (taken at 0.5 Hz, roughly the inverse of the typical time scale of the experiment), the ratio is 1.8 using a surface tension $\gamma$, initial cylinder radius $D_{0}$, and elastic modulus $G'$ of respectively, 67 mN/m \cite{Vorvolakos,Kuihua}, 300 $\mu$m and 125 Pa; hence no purely elastic instability can occur. Although the elastic sample at pH 2.5 does not exhibit BOAS, the still rather elastic state at pH 2.8 (break up dynamics shown in Fig.~\ref{F2}(b)), does show this instability. At pH 2.8 somewhat smaller relaxation time and storage modulus are found, but the calculated ratio of length scales is 2.6 which is still smaller than the limiting value of 6 reported by Mora \textit{et al.} \cite{Mora2010}. It is worthwhile noting that the shear modulus in the system of Mora \textit{et al.} is the zero frequency shear modulus. In our system, since the elastic state flows at a long time scale one would not anticipate a purely elastic instability in any case. The origin of this BOAS instability must then still be controlled by viscous rather than elastic forces, and induced by local symmetry breaking in the fluid neck \cite{Wagner2005, Deblais2018}, as for low-viscosity polymer solutions. In the experiments, BOAS indeed always start by a symmetry breaking in the filament from which a bead forms together with the formation of the elastic string. In Fig.~\ref{F3} we compare the experimentally observed growth rate of the Rayleigh-Plateau instability ${\omega_R}$ with the inverse relaxation time $\lambda_{0}^{-1}$, which is a quantitative criterion for the occurrence of a \textit{fluid} BOAS instability \cite{Wagner2005, Deblais2018}: a bead is formed until ${\omega_R} > \lambda_{0}^{-1}$. The growth rate is determined from the initial perturbation in the very beginning of regime I\cite{Wagner2005, Deblais2018}, and further details for the determination of the growth rate can be found in Supp. Fig.~8 \cite{Sup} . While for most of the HA solutions a similar growth rate is observed, the relaxation time $\lambda_{0}$ strongly increases when the pH approaches 2.5 (elastic state). This causes $\omega_{R} \sim \lambda_{0}^{-1}$, preventing a BOAS structure to occur. 

We now focus on the detailed comparison with the elastic and viscoelastic theories \cite{Turkoz2018, Eggers2020} for the shape of the interface. Figure \ref{F4} shows the interface profiles during the filament thinning and near the main drop attached to the plate, that forms due to capillary instability. As time progresses, the neck region connecting the thread with the drop becomes steeper and converges to a universal shape. Comparing these interface shapes to the recent elastic and viscoelastic calculations using the Oldroyd-B model for viscoelastic fluids and neo-Hookean calculations for elastic solids, we find an excellent agreement: all converge to the same universal self-similar solution profile (black line) showing that indeed the elastic and viscoelastic instabilities have the exact same signature.

\begin{figure}[t]
\includegraphics[scale=1]{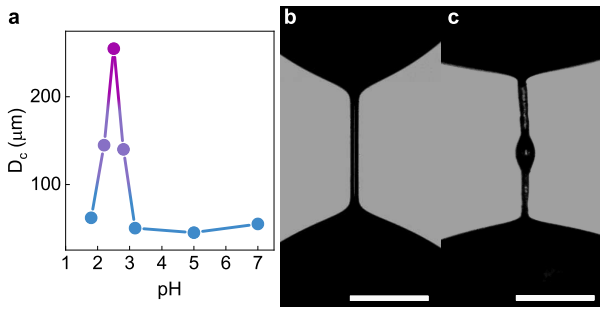}
\caption{(a) Critical neck diameter measured at the occurrence of the blistering pattern as a function of pH. Blistering is observed for a larger neck diameter when the pH approaches 2.5. Photographs of the break up dynamics of two samples of 5 mm length at pH 1.9 extended at 100 $\mu$m/s (b) and 1000 $\mu$m/s (c). In (c) a pearl instability is observed (see Supplemental video 1). The apparent roughness of the thread is consistent with polymer material while the smoothness of the pearl indicates it contains the solvent. Scale bar is 2.5 mm.}
\label{F5}
\end{figure}

Finally, we focus on the last stages of the breakup, where yet another instability is observed. Fig.~\ref{F2}(a) shows that in the last photos before the break up indicated by (iv), blistering patterns are observed at all pH values. These may superficially look somewhat similar to the BOAS instabilities, but it has been argued recently that the origin is very different, namely a partial phase separation between the polymer and the solvent that is induced by the stretching \cite{Eggers2014, Deblais2018}. For dilute polymer solutions, this has been observed to happen \cite{Deblais2018} below a critical filament diameter $D_c = \sqrt{\Delta/\dot{\epsilon}}$ = $\sqrt{3\Delta\lambda_0}$, with $\dot{\epsilon}$ being the stretching rate defined as $\dot{\epsilon} =1/3\lambda_0$ and $\Delta = k_{b}T/6\pi\eta_{s}a$ \cite{Eggers2014}, with $a$ being the polymer radius and $\eta_{s}$ the viscosity of the solvent. 
This calculation compares the time scale of advection of concentration heterogeneities to their homogenization by diffusion. We observed that in our much more concentrated system, the flow-concentration coupling must be substantially more complicated than assumed by the above calculation. When we put the numbers of our system in this calculation we find for pH 2.5 that the critical radius should be on the order of 10 nm, whereas the data shows the emergence of blistering at a radius that is approximately three orders of magnitude larger (see Fig.~\ref{F5}(a)). This therefore necessitates further theoretical study. The fact that the idea of flow-induced phase separation is nevertheless the correct one follows from the qualitative observation that the pearling instability appears sooner when stretching the filament faster, \textit{i.e}. at higher $\dot{\epsilon}$. One typical example is Fig.~\ref{F5}(b,c) where we show that, provided the stretching is fast enough, the phase separation is even directly visible in our experiments. The blisters remain smooth and contain mostly solvent, whereas the filament itself becomes rough because it is highly concentrated in polymer. The roughness of the filament is not an evaporation effect: the emergence of the blisters on a solid string is reversible (supplemental video 1). 

In conclusion, we have used a material with tunable viscoelasticity to bridge the gap between very fluid, viscoelastic and elastic polymeric materials. We have studied the process of filament formation and its ensuing instabilities, and have developed a quantitative understanding of the rich dynamics of the different processes that occur. Perhaps surprisingly, viscoelastic fluid mechanics and solid mechanics lead to a unified description of the dynamics, that is borne out in notably the interface profiles of the BOAS structure. These are first observed to have a shape given by both elastic and viscoelastic theory, that at later times converges to the universal interface profile, and agrees excellently with the theoretical prediction for this profile. These results open the way to a better understanding of the correspondence between fluid-mechanical theories that operate in a Lagrangian frame of reference and elasticity theory that uses the Eulerian reference frame. In addition, the understanding and control of the surface instabilities can have important repercussions for both fibre spinning of regular polymers \cite{Naraghi2007,Keshavarz2020}, and a wealth of instabilities observed for biopolymer systems similar to the one studied here \cite{Chaudhuri2016a,Lundahl2018}.\\

We thank J. Eggers and J. Snoeijer for helpful discussions and R. Bosschaert for performing pilot experiments. A.D. acknowledges the funding from the Horizon 2020 program under the Individual Marie Skłodowska-Curie fellowship number 798455. The work of F.B G.H.K., and D.B. is part of the IPP project Hybdrid Soft Materials carried out under an agreement with Unilever Research and Development B.V. and NWO.

\clearpage
\newpage
\renewcommand{\figurename}{Sup. Fig.}
\setcounter{figure}{0} 

\section*{Supplemental Material\\ for\\``Capillary Thinning of Elastic and Viscoelastic Thread: from Elastocapillarity to Phase Separation"}

This section provides supporting figures to accompany the main text; it gives more details on the experimental methods, the oscillatory shear measurements, the breakup dynamics of all the solutions at different pH, the shear and extensional rheology behaviour in Regime I, the break up dynamics of the fluid state and the extensional rates for both the `fluid' and `elastic' states extracted from the break up dynamics. 

\subsection*{Supplementary Note 1: Rheology measurements}

Rheology measurements were performed with a stress-controlled rheometer (Anton Paar MCR 302), equipped with a cone plate geometry with a diameter of 50 mm and cone angle of 1$^{\degree}$. The experiments were performed at a gap size of 52 $\mu$m and at a temperature of 22 $^{\degree}$C set by a Peltier system. A humidity chamber around the geometry allow us to suppress evaporation during the whole measurement. Samples are left to equilibrate for 5 days to reach homogeneity and were loaded in the rheometer geometry using a spatula. After (thermal) equilibration measurements were performed. The elastic and viscous shear moduli were probed by performing oscillatory shear measurements at an oscillation frequency of 0.5 Hz and a strain amplitude of 0.5\%, which is well within the linear viscoelastic regime for HA (see SUPPL. FIG.1.). The average reported is the representative of at least three independent measurements. The steady shear experiments were performed by carrying out a shear rate sweep from 1 $\cdot$ 10$^{-2}$ to 1000 s$^{-1}$. The flow curves were fitted to power law following $\sigma = K\gamma^{\alpha}$, with $K$ the flow consistency index. The reported results are averages of at least three measurements for each pH. For each sample, the pH was measured using a pH meter (Hanna Instruments).

\begin{figure*}
\includegraphics[scale=1]{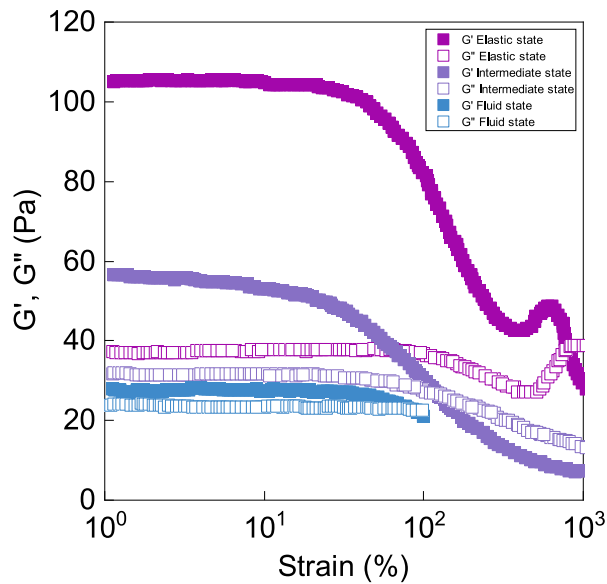}
\caption{Strain sweeps for the fluid state, intermediate state and elastic state. Further oscillatory shear measurements are well performed within the linear viscoelastic regime of HA.}
\label{SF7}
\end{figure*}

\begin{figure*}
\includegraphics[scale=1]{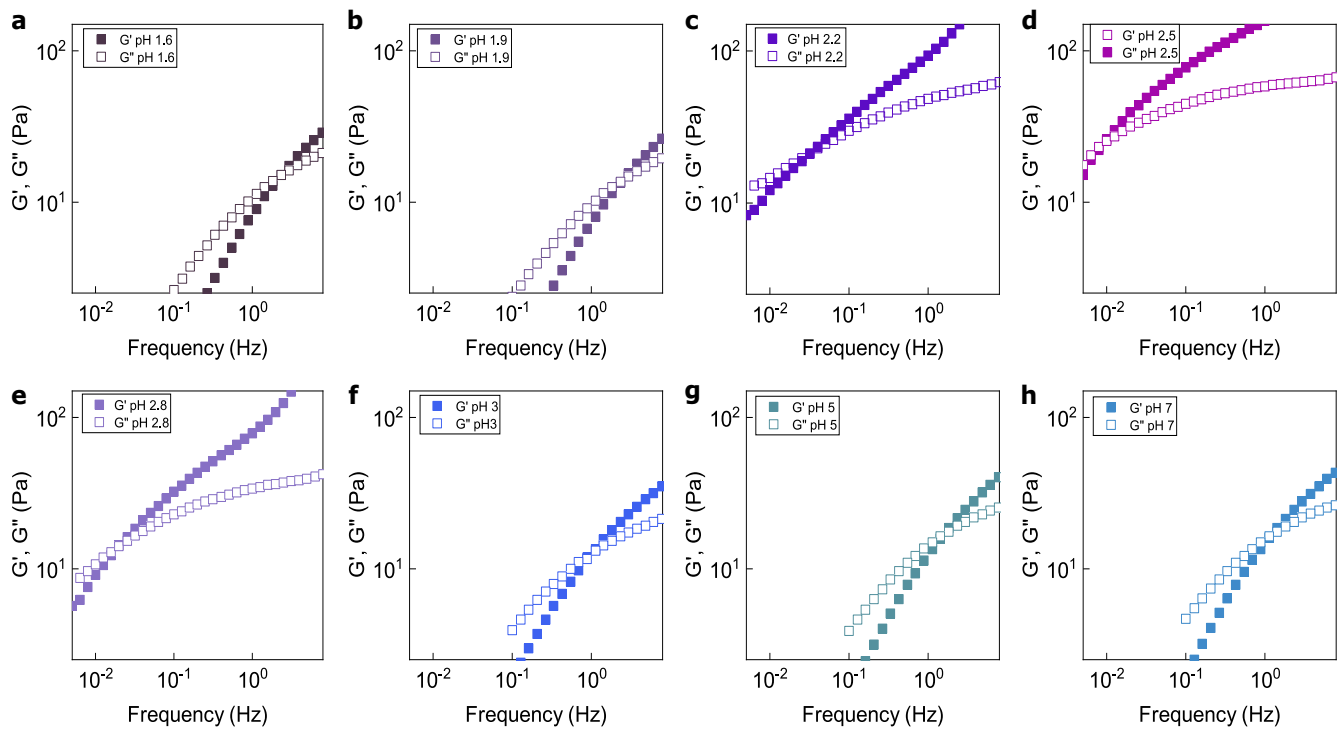}
\caption{(a-h) Oscillatory shear measurements of all pH, showing that the viscoelastic behaviour of HA strongly depends on the pH. The elastic and viscous moduli at f = 0.5 Hz and a strain amplitude of 0.5\%. For the measurements at the pH of 2.2, 2.8 and 2.5, measurements needed to be performed at lower frequencies to observe the cross over point between G' and G".} 
\label{SF4}
\end{figure*}

\begin{figure*}
\includegraphics[scale=1]{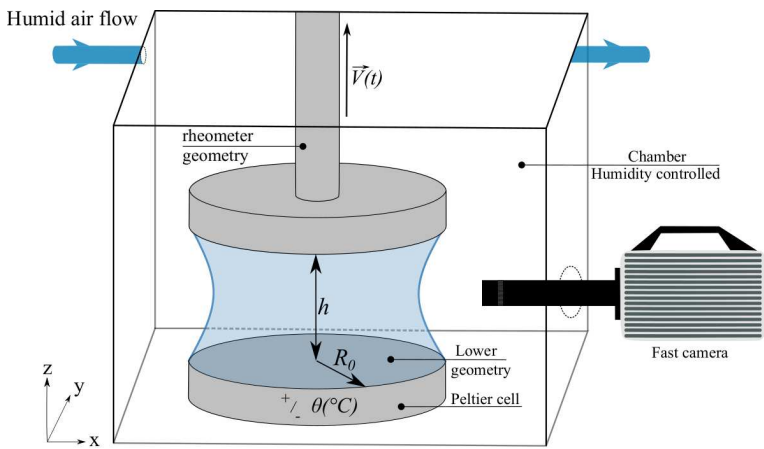}
\caption{Sketch of experimental set up (not to scale) to create a purely extensional flow to study the extensional thinning and destabilization of HA filaments. A rheometer (Anton paar, MCR 302) was used as the building block of the device. A rheometer geometry plate with a diameter of 5 mm was used, the lower plate having the same diameter. The upper plate can be pulled vertically at a constant velocity until the capillary bridge breaks. The Peltier cell allows us to impose the temperature of the sample during the elongational process at a constant speed. The evolution of the liquid bridge is recorded with a fast camera (Phantom V7) allowing frame rates up to 10.000 frames per second. The camera is equipped with a microscope tube lens, with an objective up to 12x magnification (Navitar) and a spatial resolution of 3 $\mu$m per pixel. The whole setup is placed in a chamber and is continuously flushed with humid air (80$\%$ RH) to prevent evaporation during the measurement.}
\label{SF5}
\end{figure*}

\begin{figure*}
\includegraphics[scale=1]{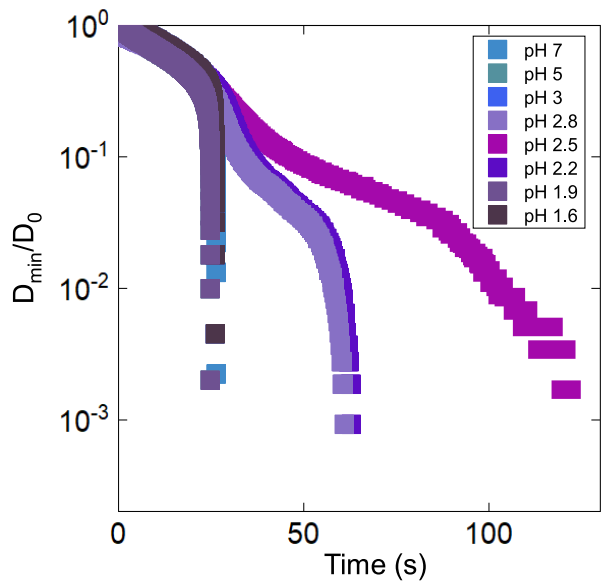}
\caption{Minimum filament diameter normalized by the initial bridge size for all pH. The dynamics of the solutions with pH 1.6, 1.9, 3, 5 and 7 are similar and there for categorized as `fluid' state in the main text. The dynamics of of pH 2.2 and 2.8 are also similar and categorized as the `intermediate' state. The solution at pH 2.5 shows the most elastic behaviour and is called the ` elastic' state.}
\label{SF6}
\end{figure*}

\begin{figure*}
\includegraphics[scale=1]{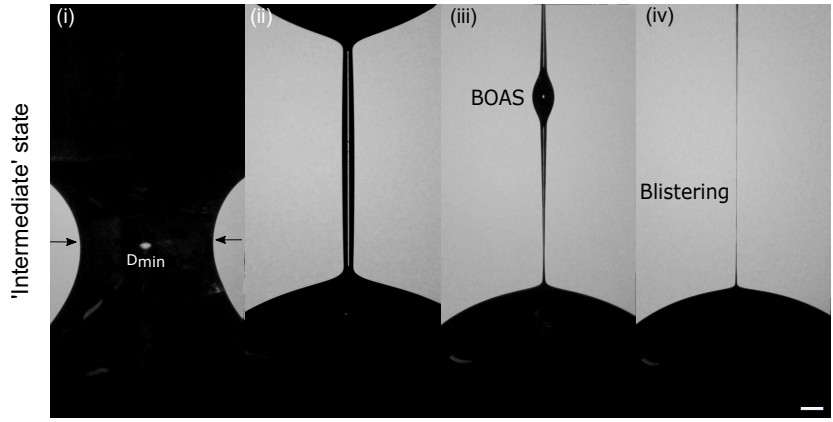}
\caption{Photographs of the breakup neck dynamics of the `intermediate' state at pH 2.8. The numbers in the pictures correspond to the different regimes; (i) the power law fluid regime (I), (ii) the exponential thinning regime (II) where a slender filament and symmetry breaking occurs, (iii) and (iv) show the BOAS and blistering instabilities respectively, which occur in the regime where the extensional viscosity is saturated (III).} 
\label{SF1}
\end{figure*}

\begin{figure*}
\includegraphics[scale=1]{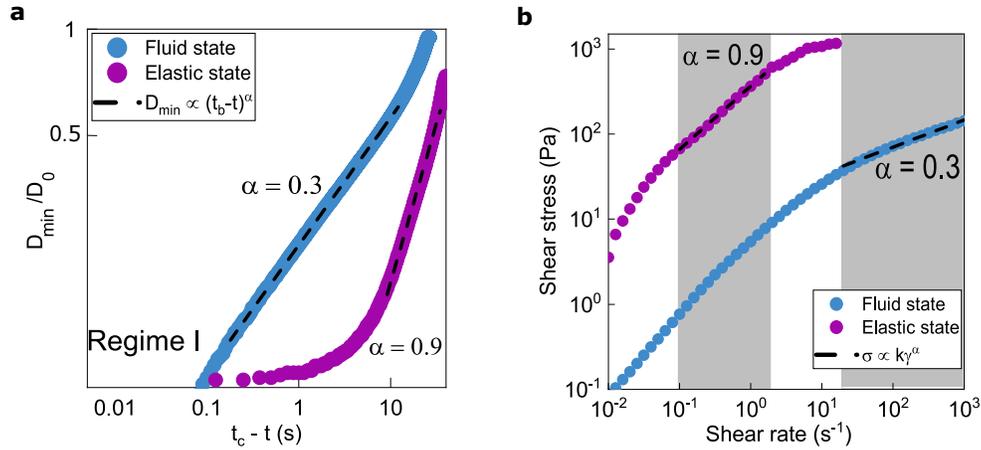}
\caption{(a) Normalized minimum neck diameter $D_{min}/D_{0}$ as a function of time $t_{b}-t$, where $t_{b}$ is the breakup time. The first part of the breakup dynamics in Regime I (main text) is plotted. The neck thinning is similar to that of a power-law fluid. This regime (I) (see main text) follows $D_{min} = (t_{b}-t)^{\alpha}$, indicated by the black dashed fit, where $\alpha$ depends on the viscous or inviscid character of the fluid. The $\alpha$ values corresponding to the fluid and elastic state (blue and purple) are respectively 0.3 and 0.9. These power law exponents from the break up dynamics correspond to the power law exponents from the shear-thinning regime of the flow curve of the steady shear experiments (b) that follows $\sigma = K\gamma^{\alpha}$, with $K$ the flow consistency index. The grey coloured areas correspond to the fitted area in (a) and to the typical deformation rates of both solutions deduced from SUPPL. FIG. 3. (red)}
\label{SF2}
\end{figure*}

\begin{figure*}
\includegraphics[scale=1]{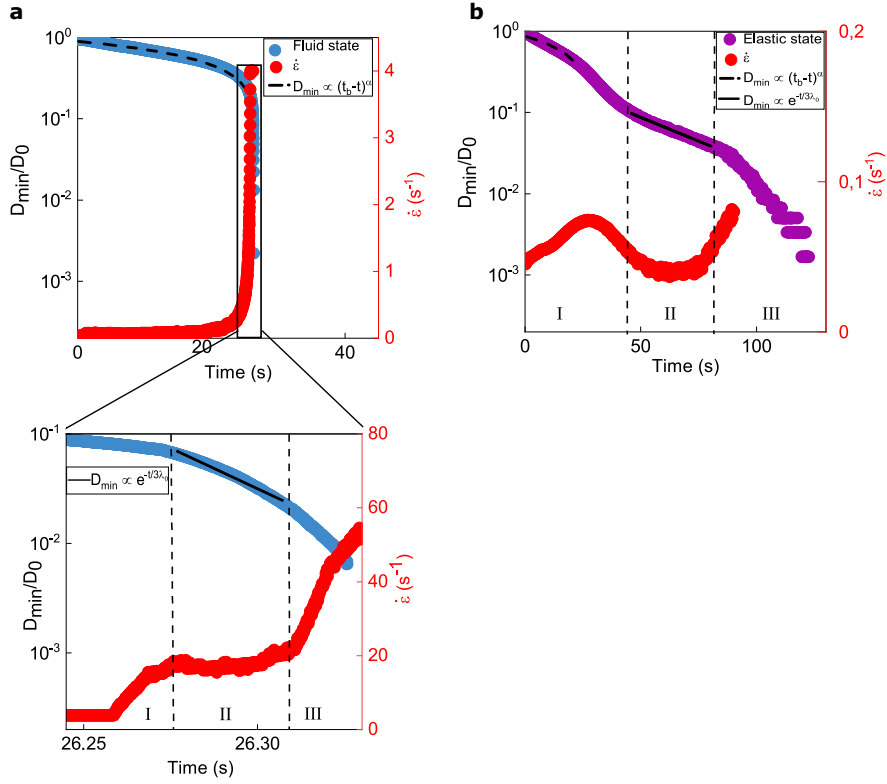}
\caption{Normalized minimum neck diameter $D_{min}/D_{0}$ and the corresponding extensional rates $\dot{\epsilon}$ as a function of time for (a) the fluid state at pH 7 with an inset where the last part of the breakup can be observed in more detail and (b) the elastic state at pH 2.5. The extensional rate is directly obtained from the evolution of the filament diameter as $\dot{{\varepsilon}}=\frac{-2}{D_{min}}\frac{dD_{min}}{dt}$. From the extensional rate the different regimes can clearly be distinguished. The extensional rate increases as the neck diameter decreases corresponding to the powerlaw regime (I). Subsequently, the extensional rate is increasing and reaching a constant value where a long and slender filament is formed corresponding to the exponential thinning regime (II). After this regime, the extensional rate keeps increasing. This leads to an even steeper exponential decay (regime III) and eventually causing the breakup of the filament. Higher values of the extensional rate are found for the fluid state as this in agreement with smaller relaxation times (of milliseconds compared to seconds in the elastic state).}
\label{SF3}
\end{figure*}

\begin{figure*}
\includegraphics[scale=1]{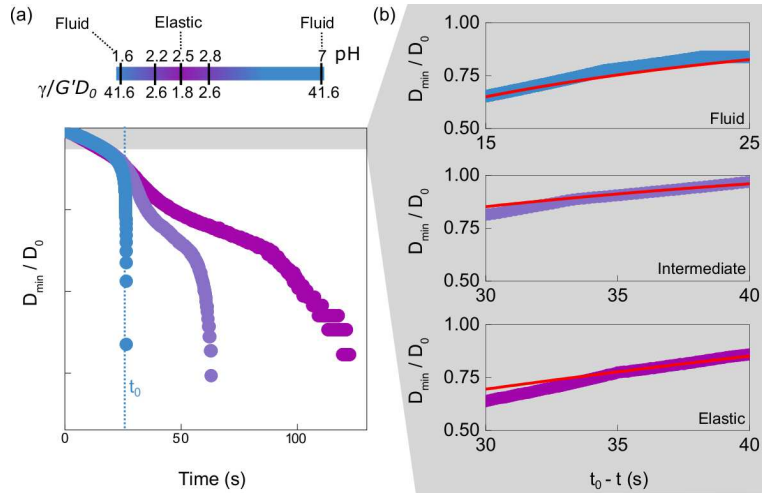}
\caption{(a) Minimum neck radius versus time in the different material state (different pH), where $t_{0}$ is the extrapolated time at which pinch off would take place without the visco-elasto-capillary regime. Same colored code as in the main text. (b) The red straight lines are exponential fits to the disturbance amplitude $1-D_{min}/D_{0}$ in the very initial times of the thinning dynamics, from which we extracted the growth rates reported in Fig. 3 of the main text.}
\label{SF8}
\end{figure*}


\begin{thebibliography}{48}%
\makeatletter
\providecommand \@ifxundefined [1]{%
 \@ifx{#1\undefined}
}%
\providecommand \@ifnum [1]{%
 \ifnum #1\expandafter \@firstoftwo
 \else \expandafter \@secondoftwo
 \fi
}%
\providecommand \@ifx [1]{%
 \ifx #1\expandafter \@firstoftwo
 \else \expandafter \@secondoftwo
 \fi
}%
\providecommand \natexlab [1]{#1}%
\providecommand \enquote  [1]{``#1''}%
\providecommand \bibnamefont  [1]{#1}%
\providecommand \bibfnamefont [1]{#1}%
\providecommand \citenamefont [1]{#1}%
\providecommand \href@noop [0]{\@secondoftwo}%
\providecommand \href [0]{\begingroup \@sanitize@url \@href}%
\providecommand \@href[1]{\@@startlink{#1}\@@href}%
\providecommand \@@href[1]{\endgroup#1\@@endlink}%
\providecommand \@sanitize@url [0]{\catcode `\\12\catcode `\$12\catcode
  `\&12\catcode `\#12\catcode `\^12\catcode `\_12\catcode `\%12\relax}%
\providecommand \@@startlink[1]{}%
\providecommand \@@endlink[0]{}%
\providecommand \url  [0]{\begingroup\@sanitize@url \@url }%
\providecommand \@url [1]{\endgroup\@href {#1}{\urlprefix }}%
\providecommand \urlprefix  [0]{URL }%
\providecommand \Eprint [0]{\href }%
\providecommand \doibase [0]{https://doi.org/}%
\providecommand \selectlanguage [0]{\@gobble}%
\providecommand \bibinfo  [0]{\@secondoftwo}%
\providecommand \bibfield  [0]{\@secondoftwo}%
\providecommand \translation [1]{[#1]}%
\providecommand \BibitemOpen [0]{}%
\providecommand \bibitemStop [0]{}%
\providecommand \bibitemNoStop [0]{.\EOS\space}%
\providecommand \EOS [0]{\spacefactor3000\relax}%
\providecommand \BibitemShut  [1]{\csname bibitem#1\endcsname}%
\let\auto@bib@innerbib\@empty
\bibitem [{\citenamefont {Larson}(1999)}]{Larson1999}%
  \BibitemOpen
  \bibfield  {author} {\bibinfo {author} {\bibfnamefont {R.~G.}\ \bibnamefont
  {Larson}},\ }\href@noop {} {\emph {\bibinfo {title} {The structure and
  rheology of complex fluids}}},\ Vol.\ \bibinfo {volume} {150}\ (\bibinfo
  {publisher} {Oxford university press New York},\ \bibinfo {year}
  {1999})\BibitemShut {NoStop}%
\bibitem [{\citenamefont {Amarouchene}\ \emph {et~al.}(2001)\citenamefont
  {Amarouchene}, \citenamefont {Bonn}, \citenamefont {Meunier},\ and\
  \citenamefont {Kellay}}]{Amarouchene2001}%
  \BibitemOpen
  \bibfield  {author} {\bibinfo {author} {\bibfnamefont {Y.}~\bibnamefont
  {Amarouchene}}, \bibinfo {author} {\bibfnamefont {D.}~\bibnamefont {Bonn}},
  \bibinfo {author} {\bibfnamefont {J.}~\bibnamefont {Meunier}},\ and\ \bibinfo
  {author} {\bibfnamefont {H.}~\bibnamefont {Kellay}},\ }\bibfield  {title}
  {\bibinfo {title} {Inhibition of the finite-time singularity during droplet
  fission of a polymeric fluid},\ }\href@noop {} {\bibfield  {journal}
  {\bibinfo  {journal} {Physical Review Letters}\ }\textbf {\bibinfo {volume}
  {86}},\ \bibinfo {pages} {3558} (\bibinfo {year} {2001})}\BibitemShut
  {NoStop}%
\bibitem [{\citenamefont {McKinley}(2005)}]{Mckinley2005}%
  \BibitemOpen
  \bibfield  {author} {\bibinfo {author} {\bibfnamefont {G.~H.}\ \bibnamefont
  {McKinley}},\ }\bibfield  {title} {\bibinfo {title} {Visco-elasto-capillary
  thinning and break-up of complex fluids},\ }\href@noop {} {\bibfield
  {journal} {\bibinfo  {journal} {Annual Rheology Reviews}\ } (\bibinfo {year}
  {2005})}\BibitemShut {NoStop}%
\bibitem [{\citenamefont {Naraghi}\ \emph {et~al.}(2007)\citenamefont
  {Naraghi}, \citenamefont {Chasiotis}, \citenamefont {Kahn}, \citenamefont
  {Wen},\ and\ \citenamefont {Dzenis}}]{Naraghi2007}%
  \BibitemOpen
  \bibfield  {author} {\bibinfo {author} {\bibfnamefont {M.}~\bibnamefont
  {Naraghi}}, \bibinfo {author} {\bibfnamefont {I.}~\bibnamefont {Chasiotis}},
  \bibinfo {author} {\bibfnamefont {H.}~\bibnamefont {Kahn}}, \bibinfo {author}
  {\bibfnamefont {Y.}~\bibnamefont {Wen}},\ and\ \bibinfo {author}
  {\bibfnamefont {Y.}~\bibnamefont {Dzenis}},\ }\bibfield  {title} {\bibinfo
  {title} {{Mechanical deformation and failure of electrospun polyacrylonitrile
  nanofibers as a function of strain rate}},\ }\href
  {http://aip.scitation.org/doi/10.1063/1.2795799} {\bibfield  {journal}
  {\bibinfo  {journal} {Applied Physics Letters}\ }\textbf {\bibinfo {volume}
  {91}},\ \bibinfo {pages} {151901} (\bibinfo {year} {2007})}\BibitemShut
  {NoStop}%
\bibitem [{\citenamefont {Keshavarz}\ \emph {et~al.}(2020)\citenamefont
  {Keshavarz}, \citenamefont {Houze}, \citenamefont {Moore}, \citenamefont
  {Koerner},\ and\ \citenamefont {McKinley}}]{Keshavarz2020}%
  \BibitemOpen
  \bibfield  {author} {\bibinfo {author} {\bibfnamefont {B.}~\bibnamefont
  {Keshavarz}}, \bibinfo {author} {\bibfnamefont {E.~C.}\ \bibnamefont
  {Houze}}, \bibinfo {author} {\bibfnamefont {J.~R.}\ \bibnamefont {Moore}},
  \bibinfo {author} {\bibfnamefont {M.~R.}\ \bibnamefont {Koerner}},\ and\
  \bibinfo {author} {\bibfnamefont {G.~H.}\ \bibnamefont {McKinley}},\
  }\bibfield  {title} {\bibinfo {title} {Rotary atomization of newtonian and
  viscoelastic liquids},\ }\href@noop {} {\bibfield  {journal} {\bibinfo
  {journal} {Physical Review Fluids}\ }\textbf {\bibinfo {volume} {5}},\
  \bibinfo {pages} {033601} (\bibinfo {year} {2020})}\BibitemShut {NoStop}%
\bibitem [{\citenamefont {Anna}\ and\ \citenamefont
  {McKinley}(2001)}]{Anna2001}%
  \BibitemOpen
  \bibfield  {author} {\bibinfo {author} {\bibfnamefont {S.~L.}\ \bibnamefont
  {Anna}}\ and\ \bibinfo {author} {\bibfnamefont {G.~H.}\ \bibnamefont
  {McKinley}},\ }\bibfield  {title} {\bibinfo {title} {{Elasto-capillary
  thinning and breakup of model elastic liquids}},\ }\href@noop {} {\bibfield
  {journal} {\bibinfo  {journal} {Journal of Rheology}\ }\textbf {\bibinfo
  {volume} {45}},\ \bibinfo {pages} {115} (\bibinfo {year} {2001})}\BibitemShut
  {NoStop}%
\bibitem [{\citenamefont {McKinley}\ and\ \citenamefont
  {Sridhar}(2002)}]{Mckinley2002}%
  \BibitemOpen
  \bibfield  {author} {\bibinfo {author} {\bibfnamefont {G.~H.}\ \bibnamefont
  {McKinley}}\ and\ \bibinfo {author} {\bibfnamefont {T.}~\bibnamefont
  {Sridhar}},\ }\bibfield  {title} {\bibinfo {title} {Filament-stretching
  rheometry of complex fluids},\ }\href@noop {} {\bibfield  {journal} {\bibinfo
   {journal} {Annual Review of Fluid Mechanics}\ }\textbf {\bibinfo {volume}
  {34}},\ \bibinfo {pages} {375} (\bibinfo {year} {2002})}\BibitemShut
  {NoStop}%
\bibitem [{\citenamefont {Furbank}\ and\ \citenamefont
  {Morris}(2004)}]{Furbank2004}%
  \BibitemOpen
  \bibfield  {author} {\bibinfo {author} {\bibfnamefont {R.~J.}\ \bibnamefont
  {Furbank}}\ and\ \bibinfo {author} {\bibfnamefont {J.~F.}\ \bibnamefont
  {Morris}},\ }\bibfield  {title} {\bibinfo {title} {An experimental study of
  particle effects on drop formation},\ }\href@noop {} {\bibfield  {journal}
  {\bibinfo  {journal} {Physics of Fluids}\ }\textbf {\bibinfo {volume} {16}},\
  \bibinfo {pages} {1777} (\bibinfo {year} {2004})}\BibitemShut {NoStop}%
\bibitem [{\citenamefont {Suryo}\ and\ \citenamefont
  {Basaran}(2006)}]{Suryo2006}%
  \BibitemOpen
  \bibfield  {author} {\bibinfo {author} {\bibfnamefont {R.}~\bibnamefont
  {Suryo}}\ and\ \bibinfo {author} {\bibfnamefont {O.~A.}\ \bibnamefont
  {Basaran}},\ }\bibfield  {title} {\bibinfo {title} {Local dynamics during
  pinch-off of liquid threads of power law fluids: Scaling analysis and
  self-similarity},\ }\href@noop {} {\bibfield  {journal} {\bibinfo  {journal}
  {Journal of non-newtonian fluid mechanics}\ }\textbf {\bibinfo {volume}
  {138}},\ \bibinfo {pages} {134} (\bibinfo {year} {2006})}\BibitemShut
  {NoStop}%
\bibitem [{\citenamefont {Smith}\ \emph {et~al.}(2010)\citenamefont {Smith},
  \citenamefont {Besseling}, \citenamefont {Cates},\ and\ \citenamefont
  {Bertola}}]{Smith2010}%
  \BibitemOpen
  \bibfield  {author} {\bibinfo {author} {\bibfnamefont {M.}~\bibnamefont
  {Smith}}, \bibinfo {author} {\bibfnamefont {R.}~\bibnamefont {Besseling}},
  \bibinfo {author} {\bibfnamefont {M.}~\bibnamefont {Cates}},\ and\ \bibinfo
  {author} {\bibfnamefont {V.}~\bibnamefont {Bertola}},\ }\bibfield  {title}
  {\bibinfo {title} {Dilatancy in the flow and fracture of stretched colloidal
  suspensions},\ }\href@noop {} {\bibfield  {journal} {\bibinfo  {journal}
  {Nature communications}\ }\textbf {\bibinfo {volume} {1}},\ \bibinfo {pages}
  {1} (\bibinfo {year} {2010})}\BibitemShut {NoStop}%
\bibitem [{\citenamefont {Miskin}\ and\ \citenamefont
  {Jaeger}(2012)}]{Miskin2012}%
  \BibitemOpen
  \bibfield  {author} {\bibinfo {author} {\bibfnamefont {M.~Z.}\ \bibnamefont
  {Miskin}}\ and\ \bibinfo {author} {\bibfnamefont {H.~M.}\ \bibnamefont
  {Jaeger}},\ }\bibfield  {title} {\bibinfo {title} {Droplet formation and
  scaling in dense suspensions},\ }\href@noop {} {\bibfield  {journal}
  {\bibinfo  {journal} {Proceedings of the National Academy of Sciences}\
  }\textbf {\bibinfo {volume} {109}},\ \bibinfo {pages} {4389} (\bibinfo {year}
  {2012})}\BibitemShut {NoStop}%
\bibitem [{\citenamefont {Huisman}\ \emph {et~al.}(2012)\citenamefont
  {Huisman}, \citenamefont {Friedman},\ and\ \citenamefont
  {Taborek}}]{Huisman2012}%
  \BibitemOpen
  \bibfield  {author} {\bibinfo {author} {\bibfnamefont {F.}~\bibnamefont
  {Huisman}}, \bibinfo {author} {\bibfnamefont {S.}~\bibnamefont {Friedman}},\
  and\ \bibinfo {author} {\bibfnamefont {P.}~\bibnamefont {Taborek}},\
  }\bibfield  {title} {\bibinfo {title} {Pinch-off dynamics in foams, emulsions
  and suspensions},\ }\href@noop {} {\bibfield  {journal} {\bibinfo  {journal}
  {Soft Matter}\ }\textbf {\bibinfo {volume} {8}},\ \bibinfo {pages} {6767}
  (\bibinfo {year} {2012})}\BibitemShut {NoStop}%
\bibitem [{\citenamefont {Eggers}(1997)}]{Eggers1997}%
  \BibitemOpen
  \bibfield  {author} {\bibinfo {author} {\bibfnamefont {J.}~\bibnamefont
  {Eggers}},\ }\bibfield  {title} {\bibinfo {title} {Nonlinear dynamics and
  breakup of free-surface flows},\ }\href@noop {} {\bibfield  {journal}
  {\bibinfo  {journal} {Reviews of modern physics}\ }\textbf {\bibinfo {volume}
  {69}},\ \bibinfo {pages} {865} (\bibinfo {year} {1997})}\BibitemShut
  {NoStop}%
\bibitem [{\citenamefont {Deblais}\ \emph
  {et~al.}(2018{\natexlab{a}})\citenamefont {Deblais}, \citenamefont {Herrada},
  \citenamefont {Hauner}, \citenamefont {Velikov}, \citenamefont {Van~Roon},
  \citenamefont {Kellay}, \citenamefont {Eggers},\ and\ \citenamefont
  {Bonn}}]{Deblais2018b}%
  \BibitemOpen
  \bibfield  {author} {\bibinfo {author} {\bibfnamefont {A.}~\bibnamefont
  {Deblais}}, \bibinfo {author} {\bibfnamefont {M.}~\bibnamefont {Herrada}},
  \bibinfo {author} {\bibfnamefont {I.}~\bibnamefont {Hauner}}, \bibinfo
  {author} {\bibfnamefont {K.}~\bibnamefont {Velikov}}, \bibinfo {author}
  {\bibfnamefont {T.}~\bibnamefont {Van~Roon}}, \bibinfo {author}
  {\bibfnamefont {H.}~\bibnamefont {Kellay}}, \bibinfo {author} {\bibfnamefont
  {J.}~\bibnamefont {Eggers}},\ and\ \bibinfo {author} {\bibfnamefont
  {D.}~\bibnamefont {Bonn}},\ }\bibfield  {title} {\bibinfo {title} {Viscous
  effects on inertial drop formation},\ }\href@noop {} {\bibfield  {journal}
  {\bibinfo  {journal} {Physical review letters}\ }\textbf {\bibinfo {volume}
  {121}},\ \bibinfo {pages} {254501} (\bibinfo {year}
  {2018}{\natexlab{a}})}\BibitemShut {NoStop}%
\bibitem [{\citenamefont {Goldin}\ \emph {et~al.}(1969)\citenamefont {Goldin},
  \citenamefont {Yerushalmi}, \citenamefont {Pfeffer},\ and\ \citenamefont
  {Shinnar}}]{Goldin1969}%
  \BibitemOpen
  \bibfield  {author} {\bibinfo {author} {\bibfnamefont {M.}~\bibnamefont
  {Goldin}}, \bibinfo {author} {\bibfnamefont {J.}~\bibnamefont {Yerushalmi}},
  \bibinfo {author} {\bibfnamefont {R.}~\bibnamefont {Pfeffer}},\ and\ \bibinfo
  {author} {\bibfnamefont {R.}~\bibnamefont {Shinnar}},\ }\bibfield  {title}
  {\bibinfo {title} {Breakup of a laminar capillary jet of a viscoelastic
  fluid},\ }\href@noop {} {\bibfield  {journal} {\bibinfo  {journal} {Journal
  of Fluid Mechanics}\ }\textbf {\bibinfo {volume} {38}},\ \bibinfo {pages}
  {689} (\bibinfo {year} {1969})}\BibitemShut {NoStop}%
\bibitem [{\citenamefont {Bazilevskii}\ \emph {et~al.}(1981)\citenamefont
  {Bazilevskii}, \citenamefont {Voronkov}, \citenamefont {Entov},\ and\
  \citenamefont {Rozhkov}}]{Bazilevskii1981}%
  \BibitemOpen
  \bibfield  {author} {\bibinfo {author} {\bibfnamefont {A.}~\bibnamefont
  {Bazilevskii}}, \bibinfo {author} {\bibfnamefont {S.}~\bibnamefont
  {Voronkov}}, \bibinfo {author} {\bibfnamefont {V.}~\bibnamefont {Entov}},\
  and\ \bibinfo {author} {\bibfnamefont {A.}~\bibnamefont {Rozhkov}},\
  }\bibfield  {title} {\bibinfo {title} {Orientational effects in the
  decomposition of streams and strands of diluted polymer solutions},\ }in\
  \href@noop {} {\emph {\bibinfo {booktitle} {Sov. Phys. Dokl}}},\
  Vol.~\bibinfo {volume} {26}\ (\bibinfo {year} {1981})\ pp.\ \bibinfo {pages}
  {333--335}\BibitemShut {NoStop}%
\bibitem [{\citenamefont {Entov}\ and\ \citenamefont
  {Yarin}(1984)}]{Entov1984}%
  \BibitemOpen
  \bibfield  {author} {\bibinfo {author} {\bibfnamefont {V.}~\bibnamefont
  {Entov}}\ and\ \bibinfo {author} {\bibfnamefont {A.}~\bibnamefont {Yarin}},\
  }\bibfield  {title} {\bibinfo {title} {Influence of elastic stresses on the
  capillary breakup of jets of dilute polymer solutions},\ }\href@noop {}
  {\bibfield  {journal} {\bibinfo  {journal} {Fluid Dynamics}\ }\textbf
  {\bibinfo {volume} {19}},\ \bibinfo {pages} {21} (\bibinfo {year}
  {1984})}\BibitemShut {NoStop}%
\bibitem [{\citenamefont {Wagner}\ \emph {et~al.}(2005)\citenamefont {Wagner},
  \citenamefont {Amarouchene}, \citenamefont {Bonn},\ and\ \citenamefont
  {Eggers}}]{Wagner2005}%
  \BibitemOpen
  \bibfield  {author} {\bibinfo {author} {\bibfnamefont {C.}~\bibnamefont
  {Wagner}}, \bibinfo {author} {\bibfnamefont {Y.}~\bibnamefont {Amarouchene}},
  \bibinfo {author} {\bibfnamefont {D.}~\bibnamefont {Bonn}},\ and\ \bibinfo
  {author} {\bibfnamefont {J.}~\bibnamefont {Eggers}},\ }\bibfield  {title}
  {\bibinfo {title} {{Droplet detachment and satellite bead formation in
  viscoelastic fluids}},\ }\href
  {https://journals.aps.org/prl/pdf/10.1103/PhysRevLett.95.164504} {\bibfield
  {journal} {\bibinfo  {journal} {Physical Review Letters}\ }\textbf {\bibinfo
  {volume} {95}},\ \bibinfo {pages} {164504} (\bibinfo {year}
  {2005})}\BibitemShut {NoStop}%
\bibitem [{\citenamefont {Bhat}\ \emph {et~al.}(2010)\citenamefont {Bhat},
  \citenamefont {Appathurai}, \citenamefont {Harris}, \citenamefont {Pasquali},
  \citenamefont {McKinley},\ and\ \citenamefont {Basaran}}]{Bhat2010}%
  \BibitemOpen
  \bibfield  {author} {\bibinfo {author} {\bibfnamefont {P.~P.}\ \bibnamefont
  {Bhat}}, \bibinfo {author} {\bibfnamefont {S.}~\bibnamefont {Appathurai}},
  \bibinfo {author} {\bibfnamefont {M.~T.}\ \bibnamefont {Harris}}, \bibinfo
  {author} {\bibfnamefont {M.}~\bibnamefont {Pasquali}}, \bibinfo {author}
  {\bibfnamefont {G.~H.}\ \bibnamefont {McKinley}},\ and\ \bibinfo {author}
  {\bibfnamefont {O.~A.}\ \bibnamefont {Basaran}},\ }\bibfield  {title}
  {\bibinfo {title} {Formation of beads-on-a-string structures during break-up
  of viscoelastic filaments},\ }\href@noop {} {\bibfield  {journal} {\bibinfo
  {journal} {Nature Physics}\ }\textbf {\bibinfo {volume} {6}},\ \bibinfo
  {pages} {625} (\bibinfo {year} {2010})}\BibitemShut {NoStop}%
\bibitem [{\citenamefont {Clasen}\ \emph {et~al.}(2006)\citenamefont {Clasen},
  \citenamefont {Eggers}, \citenamefont {Fontelos}, \citenamefont {Li},\ and\
  \citenamefont {McKinley}}]{Clasen2006}%
  \BibitemOpen
  \bibfield  {author} {\bibinfo {author} {\bibfnamefont {C.}~\bibnamefont
  {Clasen}}, \bibinfo {author} {\bibfnamefont {J.}~\bibnamefont {Eggers}},
  \bibinfo {author} {\bibfnamefont {M.~A.}\ \bibnamefont {Fontelos}}, \bibinfo
  {author} {\bibfnamefont {J.}~\bibnamefont {Li}},\ and\ \bibinfo {author}
  {\bibfnamefont {G.~H.}\ \bibnamefont {McKinley}},\ }\bibfield  {title}
  {\bibinfo {title} {The beads-on-string structure of viscoelastic threads},\
  }\href@noop {} {\bibfield  {journal} {\bibinfo  {journal} {Journal of Fluid
  Mechanics}\ }\textbf {\bibinfo {volume} {556}},\ \bibinfo {pages} {283â€“308}
  (\bibinfo {year} {2006})}\BibitemShut {NoStop}%
\bibitem [{\citenamefont {Oliveira}\ and\ \citenamefont
  {McKinley}(2005)}]{Oliveira2005}%
  \BibitemOpen
  \bibfield  {author} {\bibinfo {author} {\bibfnamefont {M.~S.}\ \bibnamefont
  {Oliveira}}\ and\ \bibinfo {author} {\bibfnamefont {G.~H.}\ \bibnamefont
  {McKinley}},\ }\bibfield  {title} {\bibinfo {title} {Iterated stretching and
  multiple beads-on-a-string phenomena in dilute solutions of highly extensible
  flexible polymers},\ }\href@noop {} {\bibfield  {journal} {\bibinfo
  {journal} {Physics of fluids}\ }\textbf {\bibinfo {volume} {17}},\ \bibinfo
  {pages} {071704} (\bibinfo {year} {2005})}\BibitemShut {NoStop}%
\bibitem [{\citenamefont {Sattler}\ \emph {et~al.}(2008)\citenamefont
  {Sattler}, \citenamefont {Wagner},\ and\ \citenamefont
  {Eggers}}]{Sattler2008}%
  \BibitemOpen
  \bibfield  {author} {\bibinfo {author} {\bibfnamefont {R.}~\bibnamefont
  {Sattler}}, \bibinfo {author} {\bibfnamefont {C.}~\bibnamefont {Wagner}},\
  and\ \bibinfo {author} {\bibfnamefont {J.}~\bibnamefont {Eggers}},\
  }\bibfield  {title} {\bibinfo {title} {Blistering pattern and formation of
  nanofibers in capillary thinning of polymer solutions},\ }\href@noop {}
  {\bibfield  {journal} {\bibinfo  {journal} {Physical Review Letters}\
  }\textbf {\bibinfo {volume} {100}},\ \bibinfo {pages} {164502} (\bibinfo
  {year} {2008})}\BibitemShut {NoStop}%
\bibitem [{\citenamefont {Sattler}\ \emph {et~al.}(2012)\citenamefont
  {Sattler}, \citenamefont {Gier}, \citenamefont {Eggers},\ and\ \citenamefont
  {Wagner}}]{Sattler2012}%
  \BibitemOpen
  \bibfield  {author} {\bibinfo {author} {\bibfnamefont {R.}~\bibnamefont
  {Sattler}}, \bibinfo {author} {\bibfnamefont {S.}~\bibnamefont {Gier}},
  \bibinfo {author} {\bibfnamefont {J.}~\bibnamefont {Eggers}},\ and\ \bibinfo
  {author} {\bibfnamefont {C.}~\bibnamefont {Wagner}},\ }\bibfield  {title}
  {\bibinfo {title} {The final stages of capillary break-up of polymer
  solutions},\ }\href@noop {} {\bibfield  {journal} {\bibinfo  {journal}
  {Physics of Fluids}\ }\textbf {\bibinfo {volume} {24}},\ \bibinfo {pages}
  {023101} (\bibinfo {year} {2012})}\BibitemShut {NoStop}%
\bibitem [{\citenamefont {Mora}\ \emph {et~al.}(2010)\citenamefont {Mora},
  \citenamefont {Phou}, \citenamefont {Fromental}, \citenamefont {Pismen},\
  and\ \citenamefont {Pomeau}}]{Mora2010}%
  \BibitemOpen
  \bibfield  {author} {\bibinfo {author} {\bibfnamefont {S.}~\bibnamefont
  {Mora}}, \bibinfo {author} {\bibfnamefont {T.}~\bibnamefont {Phou}}, \bibinfo
  {author} {\bibfnamefont {J.~M.}\ \bibnamefont {Fromental}}, \bibinfo {author}
  {\bibfnamefont {L.~M.}\ \bibnamefont {Pismen}},\ and\ \bibinfo {author}
  {\bibfnamefont {Y.}~\bibnamefont {Pomeau}},\ }\bibfield  {title} {\bibinfo
  {title} {{Capillarity driven instability of a soft solid}},\ }\href
  {https://journals.aps.org/prl/pdf/10.1103/PhysRevLett.105.214301} {\bibfield
  {journal} {\bibinfo  {journal} {Physical Review Letters}\ }\textbf {\bibinfo
  {volume} {105}},\ \bibinfo {pages} {214301} (\bibinfo {year}
  {2010})}\BibitemShut {NoStop}%
\bibitem [{\citenamefont {Snoeijer}\ \emph {et~al.}(2019)\citenamefont
  {Snoeijer}, \citenamefont {Pandey}, \citenamefont {Herrada},\ and\
  \citenamefont {Eggers}}]{Snoeijer2019}%
  \BibitemOpen
  \bibfield  {author} {\bibinfo {author} {\bibfnamefont {J.~H.}\ \bibnamefont
  {Snoeijer}}, \bibinfo {author} {\bibfnamefont {A.}~\bibnamefont {Pandey}},
  \bibinfo {author} {\bibfnamefont {M.~A.}\ \bibnamefont {Herrada}},\ and\
  \bibinfo {author} {\bibfnamefont {J.}~\bibnamefont {Eggers}},\ }\href@noop {}
  {\bibinfo {title} {The relationship between viscoelasticity and elasticity}}
  (\bibinfo {year} {2019}),\ \Eprint {https://arxiv.org/abs/arXiv:1905.12339}
  {arXiv:1905.12339} \BibitemShut {NoStop}%
\bibitem [{\citenamefont {Eggers}\ \emph {et~al.}(2020)\citenamefont {Eggers},
  \citenamefont {Herrada},\ and\ \citenamefont {Snoeijer}}]{Eggers2020}%
  \BibitemOpen
  \bibfield  {author} {\bibinfo {author} {\bibfnamefont {J.}~\bibnamefont
  {Eggers}}, \bibinfo {author} {\bibfnamefont {M.~A.}\ \bibnamefont
  {Herrada}},\ and\ \bibinfo {author} {\bibfnamefont {J.~H.}\ \bibnamefont
  {Snoeijer}},\ }\bibfield  {title} {\bibinfo {title} {Self-similar breakup of
  polymeric threads as described by the oldroyd-b model},\ }\href@noop {}
  {\bibfield  {journal} {\bibinfo  {journal} {Journal of Fluid Mechanics}\
  }\textbf {\bibinfo {volume} {887}},\ \bibinfo {pages} {A19} (\bibinfo {year}
  {2020})}\BibitemShut {NoStop}%
\bibitem [{\citenamefont {Turkoz}\ \emph {et~al.}(2018)\citenamefont {Turkoz},
  \citenamefont {Lopez-Herrera}, \citenamefont {Eggers}, \citenamefont
  {Arnold},\ and\ \citenamefont {Deike}}]{Turkoz2018}%
  \BibitemOpen
  \bibfield  {author} {\bibinfo {author} {\bibfnamefont {E.}~\bibnamefont
  {Turkoz}}, \bibinfo {author} {\bibfnamefont {J.~M.}\ \bibnamefont
  {Lopez-Herrera}}, \bibinfo {author} {\bibfnamefont {J.}~\bibnamefont
  {Eggers}}, \bibinfo {author} {\bibfnamefont {C.~B.}\ \bibnamefont {Arnold}},\
  and\ \bibinfo {author} {\bibfnamefont {L.}~\bibnamefont {Deike}},\ }\bibfield
   {title} {\bibinfo {title} {Axisymmetric simulation of viscoelastic filament
  thinning with the oldroyd-b model},\ }\href
  {https://doi.org/10.1017/jfm.2018.514} {\bibfield  {journal} {\bibinfo
  {journal} {Journal of Fluid Mechanics}\ }\textbf {\bibinfo {volume} {851}},\
  \bibinfo {pages} {R2} (\bibinfo {year} {2018})}\BibitemShut {NoStop}%
\bibitem [{\citenamefont {Giubertoni}\ \emph {et~al.}(2019)\citenamefont
  {Giubertoni}, \citenamefont {Burla}, \citenamefont {Martinez-Torres},
  \citenamefont {Dutta}, \citenamefont {Pletikapic}, \citenamefont {Pelan},
  \citenamefont {Rezus}, \citenamefont {Koenderink},\ and\ \citenamefont
  {Bakker}}]{Giubertoni2019}%
  \BibitemOpen
  \bibfield  {author} {\bibinfo {author} {\bibfnamefont {G.}~\bibnamefont
  {Giubertoni}}, \bibinfo {author} {\bibfnamefont {F.}~\bibnamefont {Burla}},
  \bibinfo {author} {\bibfnamefont {C.}~\bibnamefont {Martinez-Torres}},
  \bibinfo {author} {\bibfnamefont {B.}~\bibnamefont {Dutta}}, \bibinfo
  {author} {\bibfnamefont {G.}~\bibnamefont {Pletikapic}}, \bibinfo {author}
  {\bibfnamefont {E.}~\bibnamefont {Pelan}}, \bibinfo {author} {\bibfnamefont
  {Y.~L.~A.}\ \bibnamefont {Rezus}}, \bibinfo {author} {\bibfnamefont {G.~H.}\
  \bibnamefont {Koenderink}},\ and\ \bibinfo {author} {\bibfnamefont {H.~J.}\
  \bibnamefont {Bakker}},\ }\bibfield  {title} {\bibinfo {title} {Molecular
  origin of the elastic state of aqueous hyaluronic acid},\ }\href@noop {}
  {\bibfield  {journal} {\bibinfo  {journal} {The Journal of Physical Chemistry
  B}\ }\textbf {\bibinfo {volume} {123}},\ \bibinfo {pages} {3043} (\bibinfo
  {year} {2019})}\BibitemShut {NoStop}%
\bibitem [{\citenamefont {Burla}\ \emph {et~al.}(2019)\citenamefont {Burla},
  \citenamefont {Tauber}, \citenamefont {Dussi}, \citenamefont {van~der
  Gucht},\ and\ \citenamefont {Koenderink}}]{Burla}%
  \BibitemOpen
  \bibfield  {author} {\bibinfo {author} {\bibfnamefont {F.}~\bibnamefont
  {Burla}}, \bibinfo {author} {\bibfnamefont {J.}~\bibnamefont {Tauber}},
  \bibinfo {author} {\bibfnamefont {S.}~\bibnamefont {Dussi}}, \bibinfo
  {author} {\bibfnamefont {J.}~\bibnamefont {van~der Gucht}},\ and\ \bibinfo
  {author} {\bibfnamefont {G.~H.}\ \bibnamefont {Koenderink}},\ }\bibfield
  {title} {\bibinfo {title} {Stress management in composite biopolymer
  networks},\ }\href@noop {} {\bibfield  {journal} {\bibinfo  {journal} {Nature
  Physics}\ }\textbf {\bibinfo {volume} {15}},\ \bibinfo {pages} {549â€“553}
  (\bibinfo {year} {2019})}\BibitemShut {NoStop}%
\bibitem [{Sup()}]{Sup}%
  \BibitemOpen
  \href@noop {} {}\bibinfo {note} {See Supplemental Material [url] for more
  details on the experimental methods, the oscillatory shear measurements, the
  breakup dynamics of all the solutions at different pH, the shear and
  extensional rheology behaviour in Regime I, the break up dynamics of the
  fluid state and the extensional rates for both the `fluid' and `elastic'
  state extracted from the break up dynamics.}\BibitemShut {Stop}%
\bibitem [{\citenamefont {Campelo}\ and\ \citenamefont
  {Hern{\'{a}}ndez-Machado}(2007)}]{Campelo2007}%
  \BibitemOpen
  \bibfield  {author} {\bibinfo {author} {\bibfnamefont {F.}~\bibnamefont
  {Campelo}}\ and\ \bibinfo {author} {\bibfnamefont {A.}~\bibnamefont
  {Hern{\'{a}}ndez-Machado}},\ }\bibfield  {title} {\bibinfo {title} {{Model
  for curvature-driven pearling instability in membranes}},\ }\href
  {https://doi.org/10.1103/PhysRevLett.99.088101} {\bibfield  {journal}
  {\bibinfo  {journal} {Physical Review Letters}\ }\textbf {\bibinfo {volume}
  {99}},\ \bibinfo {pages} {088101} (\bibinfo {year} {2007})}\BibitemShut
  {NoStop}%
\bibitem [{\citenamefont {Stossel}(1993)}]{Stossel1993}%
  \BibitemOpen
  \bibfield  {author} {\bibinfo {author} {\bibfnamefont {T.}~\bibnamefont
  {Stossel}},\ }\bibfield  {title} {\bibinfo {title} {{On the crawling of
  animal cells}},\ }\href {https://doi.org/10.1126/SCIENCE.8493552} {\bibfield
  {journal} {\bibinfo  {journal} {Science}\ }\textbf {\bibinfo {volume}
  {260}},\ \bibinfo {pages} {1086} (\bibinfo {year} {1993})}\BibitemShut
  {NoStop}%
\bibitem [{\citenamefont {Lee}\ and\ \citenamefont {Mooney}(2001)}]{And2001}%
  \BibitemOpen
  \bibfield  {author} {\bibinfo {author} {\bibfnamefont {K.~Y.}\ \bibnamefont
  {Lee}}\ and\ \bibinfo {author} {\bibfnamefont {D.~J.}\ \bibnamefont
  {Mooney}},\ }\bibfield  {title} {\bibinfo {title} {{2001 Chemical
  Review-Hydrogel}},\ }\href {https://doi.org/10.1021/cr000108x} {\bibfield
  {journal} {\bibinfo  {journal} {Chemical Reviews}\ }\textbf {\bibinfo
  {volume} {101}},\ \bibinfo {pages} {1869} (\bibinfo {year}
  {2001})}\BibitemShut {NoStop}%
\bibitem [{\citenamefont {Zhao}\ \emph {et~al.}(2011)\citenamefont {Zhao},
  \citenamefont {Kim}, \citenamefont {Cezar}, \citenamefont {Huebsch},
  \citenamefont {Lee}, \citenamefont {Bouhadir},\ and\ \citenamefont
  {Mooney}}]{Zhao2011}%
  \BibitemOpen
  \bibfield  {author} {\bibinfo {author} {\bibfnamefont {X.}~\bibnamefont
  {Zhao}}, \bibinfo {author} {\bibfnamefont {J.}~\bibnamefont {Kim}}, \bibinfo
  {author} {\bibfnamefont {C.~A.}\ \bibnamefont {Cezar}}, \bibinfo {author}
  {\bibfnamefont {N.}~\bibnamefont {Huebsch}}, \bibinfo {author} {\bibfnamefont
  {K.}~\bibnamefont {Lee}}, \bibinfo {author} {\bibfnamefont {K.}~\bibnamefont
  {Bouhadir}},\ and\ \bibinfo {author} {\bibfnamefont {D.~J.}\ \bibnamefont
  {Mooney}},\ }\bibfield  {title} {\bibinfo {title} {Active scaffolds for
  on-demand drug and cell delivery},\ }\href@noop {} {\bibfield  {journal}
  {\bibinfo  {journal} {Proceedings of the National Academy of Sciences}\
  }\textbf {\bibinfo {volume} {108}},\ \bibinfo {pages} {67} (\bibinfo {year}
  {2011})}\BibitemShut {NoStop}%
\bibitem [{\citenamefont {Chaudhuri}\ \emph {et~al.}(2016)\citenamefont
  {Chaudhuri}, \citenamefont {Gu}, \citenamefont {Klumpers}, \citenamefont
  {Darnell}, \citenamefont {Bencherif}, \citenamefont {Weaver}, \citenamefont
  {Huebsch}, \citenamefont {Lee}, \citenamefont {Lippens}, \citenamefont
  {Duda},\ and\ \citenamefont {Mooney}}]{Chaudhuri2016a}%
  \BibitemOpen
  \bibfield  {author} {\bibinfo {author} {\bibfnamefont {O.}~\bibnamefont
  {Chaudhuri}}, \bibinfo {author} {\bibfnamefont {L.}~\bibnamefont {Gu}},
  \bibinfo {author} {\bibfnamefont {D.}~\bibnamefont {Klumpers}}, \bibinfo
  {author} {\bibfnamefont {M.}~\bibnamefont {Darnell}}, \bibinfo {author}
  {\bibfnamefont {S.~A.}\ \bibnamefont {Bencherif}}, \bibinfo {author}
  {\bibfnamefont {J.~C.}\ \bibnamefont {Weaver}}, \bibinfo {author}
  {\bibfnamefont {N.}~\bibnamefont {Huebsch}}, \bibinfo {author} {\bibfnamefont
  {H.~P.}\ \bibnamefont {Lee}}, \bibinfo {author} {\bibfnamefont
  {E.}~\bibnamefont {Lippens}}, \bibinfo {author} {\bibfnamefont {G.~N.}\
  \bibnamefont {Duda}},\ and\ \bibinfo {author} {\bibfnamefont {D.~J.}\
  \bibnamefont {Mooney}},\ }\bibfield  {title} {\bibinfo {title} {{Hydrogels
  with tunable stress relaxation regulate stem cell fate and activity}},\
  }\href {http://www.nature.com/articles/nmat4489} {\bibfield  {journal}
  {\bibinfo  {journal} {Nature Materials}\ }\textbf {\bibinfo {volume} {15}},\
  \bibinfo {pages} {326} (\bibinfo {year} {2016})}\BibitemShut {NoStop}%
\bibitem [{\citenamefont {McKinnon}\ \emph {et~al.}(2014)\citenamefont
  {McKinnon}, \citenamefont {Domaille}, \citenamefont {Cha},\ and\
  \citenamefont {Anseth}}]{McKinnon2014}%
  \BibitemOpen
  \bibfield  {author} {\bibinfo {author} {\bibfnamefont {D.~D.}\ \bibnamefont
  {McKinnon}}, \bibinfo {author} {\bibfnamefont {D.~W.}\ \bibnamefont
  {Domaille}}, \bibinfo {author} {\bibfnamefont {J.~N.}\ \bibnamefont {Cha}},\
  and\ \bibinfo {author} {\bibfnamefont {K.~S.}\ \bibnamefont {Anseth}},\
  }\bibfield  {title} {\bibinfo {title} {{Biophysically Defined and
  Cytocompatible Covalently Adaptable Networks as Viscoelastic 3D Cell Culture
  Systems}},\ }\href {http://doi.wiley.com/10.1002/adma.201303680} {\bibfield
  {journal} {\bibinfo  {journal} {Advanced Materials}\ }\textbf {\bibinfo
  {volume} {26}},\ \bibinfo {pages} {865} (\bibinfo {year} {2014})}\BibitemShut
  {NoStop}%
\bibitem [{\citenamefont {Ji}\ \emph {et~al.}(2006)\citenamefont {Ji},
  \citenamefont {Ghosh}, \citenamefont {Shu}, \citenamefont {Li}, \citenamefont
  {Sokolov}, \citenamefont {Prestwich}, \citenamefont {Clark},\ and\
  \citenamefont {Rafailovich}}]{Ji2006}%
  \BibitemOpen
  \bibfield  {author} {\bibinfo {author} {\bibfnamefont {Y.}~\bibnamefont
  {Ji}}, \bibinfo {author} {\bibfnamefont {K.}~\bibnamefont {Ghosh}}, \bibinfo
  {author} {\bibfnamefont {X.~Z.}\ \bibnamefont {Shu}}, \bibinfo {author}
  {\bibfnamefont {B.}~\bibnamefont {Li}}, \bibinfo {author} {\bibfnamefont
  {J.~C.}\ \bibnamefont {Sokolov}}, \bibinfo {author} {\bibfnamefont {G.~D.}\
  \bibnamefont {Prestwich}}, \bibinfo {author} {\bibfnamefont {R.~A.}\
  \bibnamefont {Clark}},\ and\ \bibinfo {author} {\bibfnamefont {M.~H.}\
  \bibnamefont {Rafailovich}},\ }\bibfield  {title} {\bibinfo {title}
  {Electrospun three-dimensional hyaluronic acid nanofibrous scaffolds},\
  }\href@noop {} {\bibfield  {journal} {\bibinfo  {journal} {Biomaterials}\
  }\textbf {\bibinfo {volume} {27}},\ \bibinfo {pages} {3782 } (\bibinfo {year}
  {2006})}\BibitemShut {NoStop}%
\bibitem [{\citenamefont {Truby}\ and\ \citenamefont
  {Lewis}(2016)}]{Truby2016}%
  \BibitemOpen
  \bibfield  {author} {\bibinfo {author} {\bibfnamefont {R.~L.}\ \bibnamefont
  {Truby}}\ and\ \bibinfo {author} {\bibfnamefont {J.~A.}\ \bibnamefont
  {Lewis}},\ }\bibfield  {title} {\bibinfo {title} {{Printing soft matter in
  three dimensions}},\ }\href {http://www.nature.com/articles/nature21003}
  {\bibfield  {journal} {\bibinfo  {journal} {Nature}\ }\textbf {\bibinfo
  {volume} {540}},\ \bibinfo {pages} {371} (\bibinfo {year}
  {2016})}\BibitemShut {NoStop}%
\bibitem [{\citenamefont {Louvet}\ \emph {et~al.}(2014)\citenamefont {Louvet},
  \citenamefont {Bonn},\ and\ \citenamefont {Kellay}}]{Louvet2014}%
  \BibitemOpen
  \bibfield  {author} {\bibinfo {author} {\bibfnamefont {N.}~\bibnamefont
  {Louvet}}, \bibinfo {author} {\bibfnamefont {D.}~\bibnamefont {Bonn}},\ and\
  \bibinfo {author} {\bibfnamefont {H.}~\bibnamefont {Kellay}},\ }\bibfield
  {title} {\bibinfo {title} {Nonuniversality in the pinch-off of yield stress
  fluids: role of nonlocal rheology},\ }\href@noop {} {\bibfield  {journal}
  {\bibinfo  {journal} {Physical review letters}\ }\textbf {\bibinfo {volume}
  {113}},\ \bibinfo {pages} {218302} (\bibinfo {year} {2014})}\BibitemShut
  {NoStop}%
\bibitem [{\citenamefont {Renardy}(2002)}]{Renardy2002}%
  \BibitemOpen
  \bibfield  {author} {\bibinfo {author} {\bibfnamefont {M.}~\bibnamefont
  {Renardy}},\ }\bibfield  {title} {\bibinfo {title} {Self-similar jet breakup
  for a generalized ptt model},\ }\href@noop {} {\bibfield  {journal} {\bibinfo
   {journal} {Journal of non-newtonian fluid mechanics}\ }\textbf {\bibinfo
  {volume} {103}},\ \bibinfo {pages} {261} (\bibinfo {year}
  {2002})}\BibitemShut {NoStop}%
\bibitem [{\citenamefont {Renardy}\ and\ \citenamefont
  {Renardy}(2004)}]{Renardy2004}%
  \BibitemOpen
  \bibfield  {author} {\bibinfo {author} {\bibfnamefont {M.}~\bibnamefont
  {Renardy}}\ and\ \bibinfo {author} {\bibfnamefont {Y.}~\bibnamefont
  {Renardy}},\ }\bibfield  {title} {\bibinfo {title} {Similarity solutions for
  breakup of jets of power law fluids},\ }\href@noop {} {\bibfield  {journal}
  {\bibinfo  {journal} {Journal of non-newtonian fluid mechanics}\ }\textbf
  {\bibinfo {volume} {122}},\ \bibinfo {pages} {303} (\bibinfo {year}
  {2004})}\BibitemShut {NoStop}%
\bibitem [{\citenamefont {Doshi}\ and\ \citenamefont
  {Basaran}(2004)}]{Doshi2004}%
  \BibitemOpen
  \bibfield  {author} {\bibinfo {author} {\bibfnamefont {P.}~\bibnamefont
  {Doshi}}\ and\ \bibinfo {author} {\bibfnamefont {O.~A.}\ \bibnamefont
  {Basaran}},\ }\bibfield  {title} {\bibinfo {title} {Self-similar pinch-off of
  power law fluids},\ }\href {https://doi.org/10.1063/1.1639015} {\bibfield
  {journal} {\bibinfo  {journal} {Physics of Fluids}\ }\textbf {\bibinfo
  {volume} {16}},\ \bibinfo {pages} {585} (\bibinfo {year} {2004})}\BibitemShut
  {NoStop}%
\bibitem [{\citenamefont {Clasen}(2010)}]{Clasen2010}%
  \BibitemOpen
  \bibfield  {author} {\bibinfo {author} {\bibfnamefont {C.}~\bibnamefont
  {Clasen}},\ }\bibfield  {title} {\bibinfo {title} {Capillary breakup
  extensional rheometry of semi-dilute polymer solutions},\ }\href@noop {}
  {\bibfield  {journal} {\bibinfo  {journal} {Korea-Australia Rheology
  Journal}\ }\textbf {\bibinfo {volume} {22}},\ \bibinfo {pages} {331}
  (\bibinfo {year} {2010})}\BibitemShut {NoStop}%
\bibitem [{\citenamefont {Vorvolakos}\ \emph {et~al.}(2014)\citenamefont
  {Vorvolakos}, \citenamefont {Coburn},\ and\ \citenamefont
  {Saylor}}]{Vorvolakos}%
  \BibitemOpen
  \bibfield  {author} {\bibinfo {author} {\bibfnamefont {K.}~\bibnamefont
  {Vorvolakos}}, \bibinfo {author} {\bibfnamefont {J.~C.}\ \bibnamefont
  {Coburn}},\ and\ \bibinfo {author} {\bibfnamefont {D.~M.}\ \bibnamefont
  {Saylor}},\ }\bibfield  {title} {\bibinfo {title} {Dynamic interfacial
  behavior of viscoelastic aqueous hyaluronic acid: effects of molecular
  weight{,} concentration and interfacial velocity},\ }\href@noop {} {\bibfield
   {journal} {\bibinfo  {journal} {Soft Matter}\ }\textbf {\bibinfo {volume}
  {10}},\ \bibinfo {pages} {2304} (\bibinfo {year} {2014})}\BibitemShut
  {NoStop}%
\bibitem [{\citenamefont {Zhang}\ \emph {et~al.}(2012)\citenamefont {Zhang},
  \citenamefont {Fan}, \citenamefont {Yan}, \citenamefont {Yu},\ and\
  \citenamefont {Mo}}]{Kuihua}%
  \BibitemOpen
  \bibfield  {author} {\bibinfo {author} {\bibfnamefont {K.}~\bibnamefont
  {Zhang}}, \bibinfo {author} {\bibfnamefont {L.}~\bibnamefont {Fan}}, \bibinfo
  {author} {\bibfnamefont {Z.}~\bibnamefont {Yan}}, \bibinfo {author}
  {\bibfnamefont {Q.}~\bibnamefont {Yu}},\ and\ \bibinfo {author}
  {\bibfnamefont {X.}~\bibnamefont {Mo}},\ }\bibfield  {title} {\bibinfo
  {title} {Electrospun biomimic nanofibrous scaffolds of silk
  fibroin/hyaluronic acid for tissue engineering},\ }\href@noop {} {\bibfield
  {journal} {\bibinfo  {journal} {Journal of Biomaterials Science, Polymer
  Edition}\ }\textbf {\bibinfo {volume} {23}},\ \bibinfo {pages} {1185}
  (\bibinfo {year} {2012})}\BibitemShut {NoStop}%
\bibitem [{\citenamefont {Deblais}\ \emph
  {et~al.}(2018{\natexlab{b}})\citenamefont {Deblais}, \citenamefont
  {Velikov},\ and\ \citenamefont {Bonn}}]{Deblais2018}%
  \BibitemOpen
  \bibfield  {author} {\bibinfo {author} {\bibfnamefont {A.}~\bibnamefont
  {Deblais}}, \bibinfo {author} {\bibfnamefont {K.~P.}\ \bibnamefont
  {Velikov}},\ and\ \bibinfo {author} {\bibfnamefont {D.}~\bibnamefont
  {Bonn}},\ }\bibfield  {title} {\bibinfo {title} {{Pearling Instabilities of a
  Viscoelastic Thread}},\ }\href
  {https://journals.aps.org/prl/pdf/10.1103/PhysRevLett.120.194501} {\bibfield
  {journal} {\bibinfo  {journal} {Physical Review Letters}\ }\textbf {\bibinfo
  {volume} {120}},\ \bibinfo {pages} {194501} (\bibinfo {year}
  {2018}{\natexlab{b}})}\BibitemShut {NoStop}%
\bibitem [{\citenamefont {Eggers}(2014)}]{Eggers2014}%
  \BibitemOpen
  \bibfield  {author} {\bibinfo {author} {\bibfnamefont {J.}~\bibnamefont
  {Eggers}},\ }\bibfield  {title} {\bibinfo {title} {Instability of a polymeric
  thread},\ }\href@noop {} {\bibfield  {journal} {\bibinfo  {journal} {Physics
  of Fluids}\ }\textbf {\bibinfo {volume} {26}},\ \bibinfo {pages} {033106}
  (\bibinfo {year} {2014})}\BibitemShut {NoStop}%
\bibitem [{\citenamefont {Lundahl}\ \emph {et~al.}(2018)\citenamefont
  {Lundahl}, \citenamefont {Berta}, \citenamefont {Ago}, \citenamefont
  {Stading},\ and\ \citenamefont {Rojas}}]{Lundahl2018}%
  \BibitemOpen
  \bibfield  {author} {\bibinfo {author} {\bibfnamefont {M.~J.}\ \bibnamefont
  {Lundahl}}, \bibinfo {author} {\bibfnamefont {M.}~\bibnamefont {Berta}},
  \bibinfo {author} {\bibfnamefont {M.}~\bibnamefont {Ago}}, \bibinfo {author}
  {\bibfnamefont {M.}~\bibnamefont {Stading}},\ and\ \bibinfo {author}
  {\bibfnamefont {O.~J.}\ \bibnamefont {Rojas}},\ }\bibfield  {title} {\bibinfo
  {title} {Shear and extensional rheology of aqueous suspensions of cellulose
  nanofibrils for biopolymer-assisted filament spinning},\ }\href@noop {}
  {\bibfield  {journal} {\bibinfo  {journal} {European Polymer Journal}\
  }\textbf {\bibinfo {volume} {109}},\ \bibinfo {pages} {367} (\bibinfo {year}
  {2018})}\BibitemShut {NoStop}%
\end{thebibliography}
\end{document}